\documentclass[aps,prl,twocolumn,showpacs,groupedaddress]{revtex4}
\usepackage{graphicx}

\begin{document}


\title{Phase Conjugation of a Quantum-Degenerate Atomic 
Fermi Beam}
\author{Chris P. Search and Pierre Meystre}
\affiliation{Optical Sciences Center, The
University of Arizona, Tucson, AZ 85721}

\date{\today}

\begin{abstract}
We discuss the possibility of phase-conjugation of an atomic 
Fermi field via nonlinear wave mixing in an ultracold gas.
It is shown that for a beam of fermions incident on an
atomic phase-conjugate mirror, a time reversed backward
propagating fermionic beam is generated similar to the case
in nonlinear optics. By adopting an operational definition
of the  phase, we show that it is possible to infer the
presence of the phase-conjugate field by the loss of the
interference pattern in an atomic interferometer.
\end{abstract}

\maketitle
In contrast with the situation for classical electromagnetic
fields, there are considerable difficulties associated with
the definition of a quantum mechanical phase operator $\hat 
\phi$ for bosonic fields \cite{Dirac,Susskind,Pegg}. This 
problem has been discussed at  great length in quantum 
optics \cite{MandelWolf}, and has recently been extended in 
the context of Bose-Einstein condensation. With none of the 
numerous attempts at formally introducing a phase operator 
being fully satisfactory, Noh, Foug{\` e}res and Mandel 
adopted instead an operational approach based on an analysis 
of what is actually measured in an experiment, namely the
relative phase between interfering fields. It can be
represented by a combination of photon counting operators
that depends on the particular experimental scheme
\cite{Noh}. A similarly operational approach has been
adopted to discuss the interference of Bose-Einstein
condensates \cite{Javanainen, Cirac}.

One fundamental difference between optical and matter-wave
fields is that the latter ones can consist of either bosons
or fermions. For fermions the question of phase of the field
is even more difficult than it is for bosons. Yet, it is
gaining considerably in relevance and timeliness in view of
the spectacular progress recently achieved toward the
experimental realization of quantum-degenerate fermionic
atomic gases, with temperatures reaching as low as $T <
0.2T_F$ where $T_F$ is the Fermi temperature
\cite{Lithium1,Dema99,Hadz02,Roat02,OHara02}. While one of
the primary goals of this work has been to observe the BCS
transition to a superfluid Fermi gas
\cite{Stoof1,Timmer1,Holland1}, future experiments will
likely use these gases for atom optics experiments for which
it will be necessary to understand what role, if any, the 
phase of the Fermi field plays. 
 
Unfortunately, one cannot define a phase operator or phase 
eigenstates for an eigenmode of the Fermi field because of 
the Pauli exclusion principle: since number and phase 
are canonically conjugate variables, a 
well-defined phase requires a large uncertainty in particle 
number. Indeed, a straightforward generalization to fermions 
of the various phase operators and phase states discussed for
bosonic fields \cite{Dirac,Susskind,Pegg} leads to mathematically ill-defined results.
Therefore, it would appear that if phase is to have any
meaning in Fermi systems, it must be associated with a
multimode cooperative effect such as the order parameter,
$\Delta$, for a superfluid Fermi gas.

The goal of this Letter is to show that despite the apparent
difficulties associated with the concept of phase for
fermionic fields, it is possible to introduce it in an 
operational way. As a previous indication that this might be
possible, we recall that it is possible in principle
to operate atom interferometers with quantum-degenerate
fermionic beams, thereby measuring the relative phase of the
partial beams \cite{Mach-Zehnder,Yurke}. Here, we go one
step further and show that it is possible to phase conjugate
a fermionic beam, so that its evolution is ``time
reversed''. This is clear evidence that from an operational
point-of-view, the phase of a fermionic beam is a perfectly
appropriate concept. To avoid any possible confusion, we
emphasize at the outset that the phase under consideration
is {\em not} the phase associated with the order parameter
$\Delta$ of a Fermi system undergoing a BCS superfluid
transition but rather the phase associated with each
eigenmode of the Fermi field.   

Optical phase conjugation has been an active area of 
research in nonlinear optics for several decades 
\cite{PC_books}. In optical phase conjugation, an incident 
signal field interacts with a pump field inside a nonlinear 
medium to generate an idler field that is the time-reversed 
state of the signal field. This process can occur via 
three-wave mixing in a $\chi^{(2)}$ medium or by four-wave 
mixing in a $\chi^{(3)}$ medium \cite{Yariv}. In the context 
of classical optics, phase conjugation can be used to
correct the phase aberrations incurred by the signal field
while in quantum optics, phase conjugation via four-wave
mixing can lead to the generation of squeezed states
\cite{yuen_shapiro}.

Four-wave mixing in normal Fermi gases has been demonstrated
theoretically in Refs. \cite{Kett01,Moor01,Vill01,4wm1,4wm2}. They show that 
four-wave mixing could be interpreted in term of Bragg 
scattering off of density modulations. However, 
phase-conjugation via four-wave mixing requires a different 
configuration that necessitates the use of a superfluid gas.
Specifically, we consider two counterpropagating 
beams of atomic fermions interacting with an atomic 
phase-conjugate mirror (PCM) 
see Fig. 1. This ``mirror'' is formed by a degenerate 
Fermi gas of alkali atoms at zero temperature confined in 
the region $0\leq z \leq L$, with equal numbers of atoms, 
$N_F$, occupying two hyperfine states that we refer to as 
spin up ($\uparrow$) and spin down ($\downarrow$). The 
confinement in the $z$-direction is provided by a trapping 
potential $U_0 < 0$ that is constant in the region 
$0\lesssim z \lesssim L$ except for a small interval $\delta 
z\ll L$ around $z=0$ and $L$ where it goes smoothly to $0$.
The extent of the gas in the $x$ and $y$ directions is taken
to be infinite so that we can treat the gas as being
spatially homogeneous with density $n_F=N_F/V$ and volume
$V$.

The spin up and down atoms interact via an
attractive two-body interaction characterized by the $s$-wave
scattering length $a<0$. The existence of an attractive
interaction leads to an instability in the normal state
of the Fermi gas that results in the BCS transition to a
superfluid state characterized by the nonzero order
parameter
$\Delta$ \cite{kittel},
\begin{equation}
\Delta=-g\sum_{\bf k}\langle {\hat c}_{-{\bf k}\downarrow}
{\hat c}_{{\bf k}\uparrow} \rangle
\end{equation}
where $g=4\pi\hbar|a|/Vm$ and ${\hat c}_{{\bf k}\sigma}$ is
the annihilation operator for a fermion of momentum
$\hbar{\bf k}$ and spin
$\sigma$. The phase-conjugate mirror is therefore described
by the linearized Hamiltonian in the region
region $0<z<L$,
\begin{eqnarray}
H&=&\sum_{\bf k}\left[ \hbar\omega_{\bf k}\left( {\hat
c}^{\dagger}_{{\bf k} \uparrow} {\hat
c}_{{\bf k} \uparrow} + {\hat c}^{\dagger}_{-{\bf k}
\downarrow} {\hat
c}_{-{\bf k} \downarrow} \right) \right . \nonumber \\
&+& \left . \hbar\Delta \left( {\hat c}^{\dagger}_{{\bf
k}\uparrow} {\hat c}^{\dagger}_{-{\bf k}\downarrow} + {\hat
c}_{-{\bf k}\downarrow}{\hat c}_{{\bf k}\uparrow}\right)
\right] \label{H}
\end{eqnarray}
where $\omega_{\bf k}=\hbar k^2/2m-\hbar k_F^2/2m$ and
$k_F=(6\pi^2 n_F)^{1/3}\gg L^{-1}$ is the Fermi wave number
of the gas. For
convenience we take $\Delta$ to be real.

Before proceeding we note that the PCM may also be produced
by a three-wave mixing process that couples the fermions to
a molecular condensate as in \cite{Timmer1},
in which case $\Delta$ in Eq. (\ref{H}) is replaced by the
expectation value of the molecular field. However, unlike
four-wave mixing, phase-matching is no longer automatically
satisfied in this case \cite{Yariv}.

The Hamiltonian $H$ may be diagonalized by the Bogoliubov
transformation
\begin{eqnarray}
{\hat \alpha}_{{\bf k}\uparrow}=\cos(\theta_{\bf k}/2){\hat
c}_{{\bf k}\uparrow}
-\sin(\theta_{\bf k}/2){\hat c}^{\dagger}_{-{\bf
k}\downarrow}, \label{Bogo1} \\
{\hat \alpha}^{\dagger}_{-{\bf
k}\downarrow}=\cos(\theta_{\bf k}/2){\hat
c}^{\dagger}_{-{\bf k}\downarrow}+\sin(\theta_{\bf
k}/2){\hat c}_{{\bf
k}\uparrow}, \label{Bogo2}
\end{eqnarray}
where $\alpha_{{\bf k}\sigma}$ is an annihilation operator
for a quasiparticle in the gas with energy $\hbar\zeta_{\bf
k}=\hbar\sqrt{\omega_k^2+\Delta^2}$ and
$\tan\theta_{\bf k}=|\Delta|/\omega_k$.

\begin{figure}
\includegraphics*[width=8cm,height=4cm]{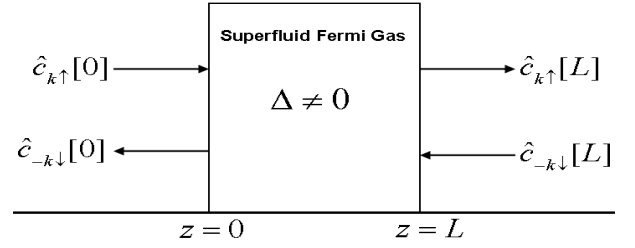}
\caption{Schematic diagram of input-output relations for
fields incident on
superfluid Fermi gas.}
\label{fig2}
\end{figure}

The cw beam impinging the PCM at $z=0$
consists of spin-up fermions with momenta $\hbar {\bf
k}=\hbar k{\bf \hat{z}}$, $k>0$.  Similarly, at $z=L$
spin-down fermions are incident on the gas with
momenta $\hbar {\bf k}=-\hbar k{\bf \hat{z}}$, $k>0$. The
beams are considered to be sufficiently well collimated in
the $x$ and $y$ directions that they can be treated as
one-dimensional \cite{Yurke}. The number of atoms in these
beams, $N_B$, satisfy $N_B\ll N_F$ so that the superfluid
gas forming the PCM can be treated as undepleted.

Now consider a fermion initially located at $z<0$ and
described as a wave-packet with average momentum $\hbar
k_0$, $\psi(z,t)\sim \psi(z-v_{k_0}(t-t_0),0)$ where
$v_{k_0}=\hbar k_0/m$ is the group velocity. When the atom
enters the superfluid gas it experiences the same confining
potential as the trapped atoms, as well as an attractive
Hartree potential, $-gn_F$, \cite{note1}
and propagates with the new group velocity
$\bar{v}_{k_0}=\hbar \bar{k}(k_0)/m$ where
\[
\bar{k}(k_0)= \sqrt{k_0^2+2m(|U_0|+gn_F)/\hbar^2},
\]
reaching the other mirror surface after a time
$\tau_{k_0}=L/\bar{v}_{k_0}$. Similarly, a
wave-packet at $z=L$ with mean momentum $-\hbar k_0$ takes a
time $\tau_{k_0}$ to reach $z=0$. The {\em input} fields are
then related to the initial conditions, ${\hat c}_{k
\uparrow}[z=0]={\hat c}_{\bar{k}(k) \uparrow}(t=0)$ and
${\hat c}_{-k \downarrow}[z=L]={\hat c}_{-\bar{k}(k)
\downarrow}(t=0)$. (To simplify the notation we define
${\hat c}_{k{\bf {\hat z}}\sigma}={\hat c}_{k \sigma}$.) The
output states, ${\hat c}_{k
\uparrow}[L]$ and ${\hat c}_{-k \downarrow}[0]$, are then
obtained by integrating the equations of motion from $t=0$
to $\tau_k$.

By using the Bogoliubov transformation
(\ref{Bogo1}-\ref{Bogo2}) and the solution of the equations
of motion for the quasiparticles,
${\hat \alpha}_{k \sigma}={\hat \alpha}_{k
\sigma}(0)\exp[-i\zeta_kt]$,
one readily obtains the output states in terms of the input
states as
\begin{eqnarray}
{\hat c}_{k \uparrow}[L]&=&T_{\bar{k}(k)}{\hat c}_{k
\uparrow}[0]+R_{\bar{k}(k)}{\hat
c}^{\dagger}_{-k \downarrow}[L] \label{in-out1} \\
{\hat c}_{-k \downarrow}[0]&=&T_{\bar{k}(k)}{\hat c}_{-k
\downarrow}[L]+R_{\bar{k}(k)}^*{\hat
c}^{\dagger}_{k \uparrow}[0] \label{in-out2}
\end{eqnarray}
where
\begin{eqnarray}
T_k&=&\cos(\zeta_kmL/\hbar k)-i\cos\theta_k
\sin(\zeta_kmL/\hbar k), \\
R_k&=&i\sin\theta_k \sin(\zeta_k mL/\hbar k).
\end{eqnarray}
Phase  conjugation only occurs when $R_k\neq 0$, the output
states being then a superposition of the transmitted input
state plus its time-reversed state, ${\hat c}^{\dagger}_{-k
\downarrow}[0]=\mathcal{T} {\hat c}_{k \uparrow}[0]
\mathcal{T}^{-1}$ where $\mathcal{T}$ is the time-reversal
operator. Note that $R_k=0$ in the absence of a superfluid
state, $\Delta=0$, so that the existence of this state is
essential for the operation of the PCM. Just as is normally
the case in optics, the phase-conjugate mirror has a finite
bandwidth, since $\sin\theta_k=|\Delta|/\zeta_k$ is only
different from zero in the interval $\delta k\approx
|\Delta|m/\hbar k_F$ around $k_F$. The phase-conjugate
signal is therefore optimized by using an input state with
average momentum $\hbar k_0$ and
bandwidth $\Delta k$ such that $\bar{k}(k_0)=k_F$ and
$|\bar{k}(k_0+\Delta k)-\bar{k}(k_0-\Delta k)|<\delta k$. In
this case one can make $|R_{k_F}|=1$ for $L=(2j+1)\pi\hbar
k_F/2m|\Delta|$ for $j=0,1,2,...$.

An important difference from the optical case \cite{Yariv}
is that since $|T_k|^2+|R_k|^2=1$, there is no amplification
of the individual modes of the fermion beams. This is in
stark contrast to the case of bosons where one has instead
$|T_k|^2-|R_k|^2=1$ in order to preserve the commutation
relations and the transmitted field is always
amplified since $|T_k|\geq 1$. The lack of amplification for
a single mode of the fermion field is a necessary
consequence of the Pauli exclusion principle. Note, however,
that the total number of fermions in the output beams can be
amplified. To see this, we take for definiteness the input
state of the fermion beams to be
\begin{equation}
|\Psi\rangle=\prod_{|k-k_0|\leq \Delta
k}\hat{c}^{\dagger}_{k\uparrow}[0]|0\rangle
\label{psi1}
\end{equation}
with $k_0>\Delta k>0$. This corresponds to a beam of spin-up fermions with momenta
centered around $\hbar k_0$ incident from $z<0$ with no
atoms incident from $z>L$. The occupation numbers for this
state are $n_k=\langle \hat{c}^{\dagger}_{k
\uparrow}[0]\hat{c}_{k
\uparrow}[0] \rangle$. Defining the total number operators for
the input and output fields as
$\hat{N}^{(\rm{in/out})}_{\uparrow}=\sum_{k>0}\hat{c}^{\dagger}_{k
\uparrow}[0/L]\hat{c}_{k \uparrow}[0/L]$ and
$\hat{N}^{({\rm
in/out})}_{\downarrow}=\sum_{k>0}\hat{c}^{\dagger}_{-k
\downarrow}[L/0]\hat{c}_{-k \downarrow}[L/0]$, one finds
that their expectation values are,
\begin{eqnarray}
\langle \hat{N}^{{\rm
out}}_{\uparrow}\rangle=\langle\hat{N}^{{\rm
in}}_{\uparrow}\rangle
+\langle \hat{N}^{{\rm out}}_{\downarrow} \rangle
\label{occ1} \\
\langle \hat{N}^{{\rm out}}_{\downarrow} \rangle=\sum_k
|R_{\bar{k}(k)}|^2 (1-n_k).
\label{occ2}
\end{eqnarray}
Eqs. (\ref{occ1}) and (\ref{occ2}) show that the number of
atoms in both beams increase after having passed through the
gas. However, Eq. (\ref{occ2}) shows that the increase
results from the scattering of atoms out of the superfluid
gas and into those modes that are not occupied in the
incident beam. Consequently, only incoherent amplification
of the vacuum fluctuations occurs.

The identical increase in both beams reflects the underlying
pair creation  process given by the ${\hat
c}^{\dagger}_{{\bf k}\uparrow} {\hat c}^{\dagger}_{-{\bf
k}\downarrow}$ term in the Hamiltonian, as well as the fact
that the number difference operator, ${\hat
c}^{\dagger}_{{\bf k} \uparrow} {\hat c}_{{\bf k}
\uparrow}-{\hat c}^{\dagger}_{-{\bf k} \downarrow} {\hat
c}_{-{\bf k} \downarrow}$ commutes with the Hamiltonian.
Defining the covariance matrix as $
{\rm Cov}[\hat{N}^{{\rm out}}_{\sigma_1}\hat{N}^{{\rm
out}}_{\sigma_2}]=
\langle \hat{N}^{{\rm out}}_{\sigma_1}\hat{N}^{{\rm
out}}_{\sigma_2}\rangle -
\langle\hat{N}^{{\rm out}}_{\sigma_1}\rangle
\langle\hat{N}^{{\rm out}}_{\sigma_2}\rangle
$ we find that
\begin{equation}
{\rm Cov}[\hat{N}^{{\rm out}}_{\sigma_1}\hat{N}^{{\rm
out}}_{\sigma_2}]=\sum_k
|T_{\bar{k}(k)}|^2|R_{\bar{k}(k)}|^2 (1-n_k) \label{COV}
\end{equation}
which shows that the intra-beam number fluctuations as well
as the correlations of the beams are the same. Again, this
reflects the fact that any atom created in one beam
coincides with an atom created in the other beam so that the
fluctuations in both beams must be the same.

Eqs. (\ref{occ2}) and (\ref{COV}) show that for
$|R_{\bar{k}(k)}|=1$ one can have amplification of the
number of atoms in the output beam with no fluctuations.
Thus by using an input state in which all the $k$ states are
fully occupied {\em except} for a narrow window of width
$\Delta k$ centered at $k_0$, one can have amplified output
beams with negligible number fluctuations provided
$\bar{k}(k_0)=k_F$, $|\bar{k}(k_0+\Delta
k)-\bar{k}(k_0-\Delta k)|<\delta k$, and $L=(2j+1)\pi\hbar
k_F/2m|\Delta|$. However, since these are the same
conditions required for a finite phase conjugate signal, one
cannot simultaneously have a phase conjugate signal {\em
and} an amplified output with reduced fluctuations.

These results show that despite the lack of a well-defined
phase for fermionic fields, the phase-conjugation and
time-reversal of these fields is readily possible in
principle. In the present example, the phase-conjugate
signal provides a direct signature of the BCS
superfluid state since for a Fermi gas in the normal state,
an incident beam of fermions would not generate a backward
propagating reflected beam \cite{note1}. Detailed studies of the
reflected beam could then be used as a diagnostic tool to
study the BCS state.

To further illustrate the analogy between fermionic and
bosonic phase conjugation, we now show how phase-conjugate
beams of fermionic atoms can compensate the relative
phase accumulated between two paths of a matter-wave
interferometer when one of the mirrors is replaced by a
phase-conjugate mirror.

Fig. 2 shows a model Michelson interferometer. Here, $BS$
labels a beam splitter with transmission and reflection
amplitudes
$t$ and $r$, respectively, with $|r|^2+|t|^2=1$ and
$rt^*+r^*t=0$. We first let $M_1$ and $M_2$ be perfectly
reflecting atom mirrors while $\ell_1$ and $\ell_2$ are the
one way path lengths in the two arms of the interferometer,
and $D$ represents an atom counter. The operators
$\hat{a}_k$ and $\hat{b}_k$ denote annihilation operators
for fermions with momentum $\hbar k$ that are incident on
the two input ports of the interferometer. The input states
only involve a single spin state so we drop the spin label
for convenience. The outputs of the beam splitter are
$\hat{g}_k=r\hat{a}_k+t\hat{b}_k$ and $\hat{f}_k=t\hat{a}_k+r\hat{b}_k$.
The mirrors introduce a phase shift assumed to be the same
for both mirrors, so that it can be ignored. After
propagation through the arms and recombination at the beam
splitter, the fermion annihilation operator at the atom
counter is
\begin{equation}
\hat{d}_k=rt(e^{-i2k\ell_2}+e^{-i2k\ell_1})\hat{a}_k+(r^2e^{-i2k\ell_2}+t^2e^{-i
2k\ell_1})\hat{b}_k.
\end{equation}
Using an input state of the same form as Eq. (\ref{psi1}),
$|\Psi '\rangle=\prod_{|k-k_0|\leq \Delta
k}\hat{a}^{\dagger}_{k}|0\rangle$, we find that the number
of atoms at $D$ is
\begin{eqnarray}
&&\langle \hat{N}_D \rangle = \sum_k \langle
\hat{d}_k^{\dagger}\hat{d}_k \rangle \nonumber \\
&&= 2|r|^2|t|^2 N_B\left(1+\cos(\Delta\ell k_0)
\frac{\sin(\Delta\ell\Delta
k)}{\Delta\ell\Delta k} \right)
\end{eqnarray}
where $\Delta\ell=2(\ell_1-\ell_2)$ and $N_B$ is the number
of incident atoms. There is a discernible interference
pattern, although the broadband nature of the Fermi beam
leads to a loss of contrast due to dephasing for large path
differences, $\Delta \ell \Delta k\gg 1$. It is worth
emphasizing that even though the incident fermions are in
Fock states which have  completely random phases $\phi_k$
for each of the occupied $k$-states, this phase is the same
in both arms of the interferometer so that the 
phase difference between the arms is independent of
$\phi_k$.

\begin{figure}
\includegraphics*[width=8cm,height=5cm]{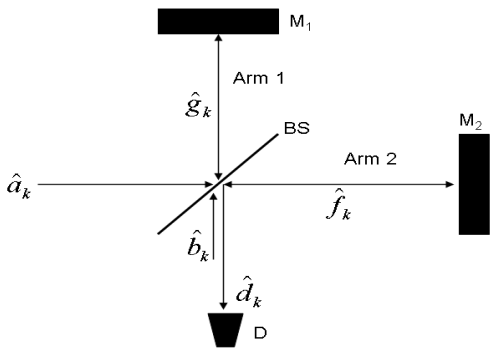}
\caption{Schematic diagram of atomic Michelson
interferometer.}
\label{fig3}
\end{figure}

We now replace the mirror $M_2$ by a phase-conjugate mirror.
The reflected output of that mirror is
\[
\hat{f}_{k}'=R_{\bar{k}(k)}^{*}\hat{f}_k^{\dagger}e^{+ik\ell_2}+T_{\bar{k}(k)} \hat{c}_k
\]
where $\hat{c}_k$ is an annihilation operator for the
fermions incident on the other side of the mirror in the
opposite spin state, see Eqs. (\ref{in-out1}-\ref{in-out2}).
The annihilation operator for fermions at the atom counter
is now
\begin{eqnarray}
\hat{d}_k&=&te^{-i2k\ell_1}(r\hat{a}_k+t\hat{b}_k)
+R^{*}_{\bar{k}(k)}r
(t^*\hat{a}_{k}^{\dagger}+r^*\hat{b}^{\dagger}_{k})\nonumber
\\
&+& rT_{\bar{k}(k)}e^{-ik\ell_2}\hat{c}_k,
\end{eqnarray}
which, again using $|\Psi'\rangle$, gives
\begin{eqnarray*}
&&\langle \hat{N}_D\rangle=\nonumber \\
&& \sum_k \left(
|r|^2|t|^2|T_{\bar{k}(k)}|^2\Theta(\Delta k-|k-k_0|)
+|r|^2|R_{\bar{k}(k)}|^2 \right)
\end{eqnarray*}
where $\Theta$ is the unit step function. In this case,
there is no detectable interference pattern: The phase in
arm 1 is $2k\ell_1+\phi_k$ while the accumulated phase after
round trip propagation in arm 2 is $-\phi_k$ due to the
phase-conjugate mirror. The phase difference between the two
arms is then $2k\ell_1+2\phi_k$ but since $\phi_k$ is a
random variable, there is no interference pattern. This
effect has also been predicted to occur in optics with
chaotic fields \cite{agarwal}.

In conclusion we have shown that phase-conjugation of Fermi
fields can be achieved using four-wave mixing in a
superfluid Fermi gas, and that it is justified to discuss
the relative phase of fermionic fields operationally. Surprisingly we have 
shown that the phase of the Fermi field can have observable effects even 
though we cannot define a Hermitian operator for the phase. In
addition, our proposed experimental setup provides an
unambiguous signature of superfluidity in the phase-conjugate mirror.

This work is supported in part by the US Office of Naval
Research, by the National Science Foundation, by the US Army
Research Office, by the National Aeronautics and Space
Administration, and by the Joint Services Optics Program.

\end{document}